\documentclass[epj,nopacs]{svjour} 
 
\usepackage{latexsym} 
\usepackage{epsfig}
\usepackage{amsmath,amssymb}
\usepackage{threeparttable}
 
\begin{document} 
 
\title{Weak form factors for semileptonic octet baryon decays\\ in the
chiral quark model} 

\author{T. Ohlsson\thanks{\email{tommy@theophys.kth.se}}, H.
Snellman\thanks{\email{snell@theophys.kth.se}}}
 
\mail{T. Ohlsson}

\institute{Theoretical Physics, Department of
Physics, Royal Institute of Technology, SE-100 44 Stockholm, Sweden} 
 
\date{Received: 9 March 1998} 

\abstract{We study the weak vector and axial-vector form factors of first- and
second-class currents for \\ the semileptonic octet baryon decays in the
spirit of the chiral quark model. Our results for the weak magnetism
form factors are consistent with the conserved vector current (CVC)
results. The induced pseudotensor form factors, which are highly model
dependent, are small. The overall performance of the chiral quark
model is quite good and in general agreement with existing
experimental data.}
 
\authorrunning{T. Ohlsson, H. Snellman}
\titlerunning{Weak form factors for semileptonic octet baryon decays in the
chiral quark model}
\maketitle

\section{Introduction}
\label{sec:intro}

The weak axial-vector form factors of the baryons have been, and still
are, an important set of parameters for the investigation of their
quark spin structure. In particular the so called ``nucleon spin
crisis'' \cite{ashm88,clos93}, as an indication of the subtle dynamics
of the quark spin polarization of the nucleons, relate the measurement
of the deep inelastic scattering (DIS) parameters to the spin polarizations
of the quarks via the baryonic axial-vector form factors.

The analysis from experiments of these form factors are normally
performed assuming that the second-class form factors are negligible.
However, for strangeness- \\ changing currents between unequal mass
states, the SU(3) symmetry breaking may induce non-negligible
second- \\ class currents. At least one experiment have reported
substantial such currents \cite{hsue88}.

Since the axial-vector form factors are used for extracting the quark
spin content of baryons, their exact values are of importance. Also
when one wants to compare the axial-vector form factors with the
Cabibbo theory, or with model calculations, it is important not to
have a mixture of first- and second-class form factors to deal with.

Pending further experiments, it is of interest to estimate these form
factors theoretically. This has been done earlier in a relativistic
quark model \cite{kell74}, in the MIT bag model
\cite{dono82,eegl85,eegh86,lies87}, and in the MIT bag model with
one-gluon QCD corrections \cite{cars88}. Recently, there has also been
a calculation within the chiral quark-soliton model \cite{kimp97}.
Unfortunately, previous results do not mutually agree on the size of
the $\Delta S = 1$ second-class form factors, and in fact not even on
the sign of them. This probably indicates that they are model
dependent.

In this paper, we estimate all six vector and axial-vector form
factors $f_i$ and $g_i$, where $i=1,2,3$, (defined in
Sect.~\ref{sec:form} below) in the spirit of the chiral quark model
($\chi$QM) \cite{mano84,eich92,chen95} to linear order in the SU(3)
symmetry breaking masses. This is of interest since the $\chi$QM gives
a fair description of the magnetic moments of the baryons, and can be
used to calculate the axial-vector form factors of the baryons in a
way that substantially deviates from the non-relativistic quark model
(NQM) due to the depolarization of the quark spins in the $\chi$QM by
the Goldstone bosons (GBs). Our estimates are made in the same
approximation as those of the magnetic moments of the baryons, treated
earlier in the literature \cite{chen952,lind97,lind98}.

Besides the axial-vector form factor $g_A \equiv g_1/f_1$ our study
will focus on the ratio $f_2/f_1$ for the vector current and the
corresponding ratio $g_2/g_1$ for the axial-vector current. In the
$\chi$QM, the ratio $g_2/g_1$ is dependent only on the mass
parameters. As shown by Donoghue and Holstein \cite{dono82}, the form
factor $g_{2}$ is essentially the axial dipole moment, which is
inversely proportional to the quark masses.

In our study, we find that the second-class form factors are small and
highly model dependent. It seems therefore even more important to
measure them, in order to find out more about the detailed dynamics of
the baryons. It makes it also possible to directly compare the
measured axial-vector form factors with the theoretical ones.

The outline of the paper is as follows. In Sect.~\ref{sec:form} we
describe the formalism and present the formulas for the first- and
second-class form factors. In Sect.~\ref{sec:numres} we make
estimates of the form factors in the $\chi$QM. We end this section
with a discussion about our results compared to other models and
experiments. Finally, in Sect.~\ref{sec:SC} we present a summary
and our conclusions.

\newpage

\section{Formalism}
\label{sec:form}

\subsection{The weak form factors}

The transition matrix element ${\cal M}_{B \to B' l^- \bar{\nu}_l}$
for the decay $B \to B' + l^- + \bar{\nu}_l$ ($q \to q' + l^- +
\bar{\nu}_l$), is given by
\begin{equation}
{\cal M}_{B \to B' l^- \bar{\nu}_l} = \frac{G}{\sqrt{2}} V_{qq'}
\langle B'(p') \vert J_{\rm weak}^\mu \vert B(p) \rangle L_\mu,
\end{equation}
where $G$ is the Fermi coupling constant, $V_{qq'}$ is the
$qq'$-element of the Cabibbo--Kobayashi--Maskawa mixing matrix, and
$L_{\mu}$ is the leptonic current.

The hadronic weak current is
\begin{equation}
J_{\rm weak}^\mu = J_V^\mu - J_A^\mu,
\end{equation}
where $J_V^\mu$ is the vector current and $J_A^\mu$ is the
axial-vector current. The matrix element of the vector current in
momentum space of the transition $B \to B' + l^- + \bar{\nu}_l$ is
given by
\begin{eqnarray}
\langle B'(p') \vert &J_V^\mu& \vert B(p) \rangle = \bar{u}'(p') \bigg(
f_1(q^2) \gamma^\mu \\
&-& i \frac{f_2(q^2)}{M_B + M_{B'}} \sigma^{\mu \nu}
q_\nu + \frac{f_3(q^2)}{M_B + M_{B'}} q^\mu \bigg) u(p) \nonumber
\label{eq:matV}
\end{eqnarray}
and the matrix element of the axial-vector current by
\begin{eqnarray}
\langle B'(p') \vert &J_A^\mu& \vert B(p) \rangle = \bar{u}'(p') \bigg(
g_1(q^2) \gamma^\mu \gamma^5 \\
&-& i \frac{g_2(q^2)}{M_B + M_{B'}}
\sigma^{\mu \nu} q_\nu \gamma^5 + \frac{g_3(q^2)}{M_B + M_{B'}} q^\mu
\gamma^5 \bigg) u(p), \nonumber
\label{eq:matA}
\end{eqnarray}
where $M_B$ ($M_{B'}$), $p$ ($p'$), $u(p)$ ($u'(p')$), and $\vert B(p)
\rangle$ \\ ($\vert B'(p') \rangle$) are the mass, momentum, Dirac
spinor, and external bar\-yon state of the initial (final) bar\-yon $B$
($B'$), respectively, and $q = p - p'$ is the momentum transfer
\cite{dono92}. The functions $f_i(q^2), i = 1,2,3$, are the vector
current form factors and the functions $g_i(q^2), i = 1,2,3$, are the
axial-vector current form factors. The form factors are Lorentz
scalars and they contain all the information about the hadron
dynamics. $f_1$ is the {\it vector} form factor, $f_2$ is the {\it
induced tensor} form factor (or {\it weak magnetism} form factor or
{\it anomalous magnetic moment} form factor), $f_3$ is the {\it
induced scalar} form factor, $g_1$ is the {\it axial-vector} form
factor, $g_2$ is the {\it induced pseudotensor} form factor (or {\it
weak electric} form factor), and $g_3$ is the {\it induced
pseudoscalar} form factor.

Under $G$-parity, the form factor $f_2$ transforms with the same sign
as the form factor $f_1$, whereas the form factor $f_3$ has the
opposite sign, and the form factor $g_3$ transforms with the same sign
as the form factor $g_1$, whereas the form factor $g_2$ has the
opposite sign. The currents with form factors $f_3$ and $g_2$ are
therefore called {\it second-class currents}, and the others are {\it
first-class currents} \cite{wein58}.

If we use the Gordon equalities
\begin{eqnarray}
\bar{u}'(p') \sigma^{\mu\nu} q_\nu u(p) &=& i \bar{u}'(p')
\big(\left(M_B + M_{B'}\right) \gamma^\mu \nonumber \\
&-& \left(p^\mu +
{p'}^\mu\right)\big) u(p)
\end{eqnarray}
and
\begin{eqnarray}
\bar{u}'(p') \sigma^{\mu\nu} q_\nu \gamma^5 u(p) &=& i \bar{u}'(p')
\big(\left(M_{B'} - M_B\right) \gamma^\mu \gamma^5 \nonumber \\
&-& \left(p^\mu +
{p'}^\mu\right) \gamma^5\big) u(p),
\label{eq:gordon}
\end{eqnarray}
we can write (\ref{eq:matV}) and (\ref{eq:matA}) as
\begin{eqnarray}
\langle B'(p') \vert &J_V^\mu& \vert B(p) \rangle \nonumber \\
&=& \bar{u}'(p')
\Bigg( \left( f_1(q^2) + f_2(q^2) \right) \gamma^\mu \nonumber \\ &-&
\frac{f_2(q^2)}{M_B + M_{B'}} \left( p^\mu + {p'}^\mu \right) +
\frac{f_3(q^2)}{M_B + M_{B'}} q^\mu \Bigg) u(p) \nonumber \\
&&
\label{eq:matelV}
\end{eqnarray}
and
\begin{eqnarray}
\langle B'(p') \vert &J_A^\mu& \vert B(p) \rangle \nonumber \\
&=& \bar{u}'(p') \Bigg( \bigg(g_1(q^2) + \frac{M_{B'} - M_B}{M_B + M_{B'}}
g_2(q^2)\bigg) \gamma^\mu \gamma^5 \nonumber \\
&-& \frac{g_2(q^2)}{M_B + M_{B'}} \left( p^\mu + {p'}^\mu \right) \gamma^5
\nonumber \\
&+& \frac{g_3(q^2)}{M_B + M_{B'}} q^\mu \gamma^5 \Bigg) u(p). \nonumber \\
&&
\label{eq:matelA}
\end{eqnarray}

In the Breit-frame, {\it i.e.} the Lorentz-frame, where ${\bf p} = -
{\bf p}' = \frac{1}{2} {\bf q}$, we obtain in the non-relativistic
limit (${\bf q}^2 \ll M_B^2, M_{B'}^2$)
\begin{eqnarray}
\langle B' \vert J_V^0 \vert B \rangle &=& N_B N_{B'}^\ast
{\chi'}^\dagger \left[ f_1(q^2) + \frac{M_B - M_{B'}}{M_B + M_{B'}}
f_3(q^2) \right] \chi \nonumber \\ 
&& \\
\langle B' \vert J_V^i \vert B \rangle &=&
N_B N_{B'}^\ast {\chi'}^\dagger \Bigg[ \bigg( \frac{1}{M_B + M_{B'}}
f_3(q^2) \nonumber \\
&-& \frac{M_B - M_{B'}}{4 M_B M_{B'}} \left(f_1(q^2) +
f_2(q^2)\right) \bigg) q^i \nonumber \\ &+& i \epsilon^{ijk}
\frac{M_B + M_{B'}}{4 M_B M_{B'}} \left( f_1(q^2) + f_2(q^2) \right)
q^j \sigma^k \Bigg] \chi \nonumber \\
&&
\end{eqnarray}
for the vector current and
\begin{eqnarray}
\langle B' \vert J_A^0 \vert B \rangle &=& N_B N_{B'}^\ast
{\chi'}^\dagger \bigg[ \frac{M_B - M_{B'}}{4 M_B M_{B'}} \left( g_3(q^2)
- g_1(q^2) \right) \nonumber \\
&-& \frac{1}{M_B + M_{B'}} g_2(q^2) \bigg]
{\mbox{\boldmath$\sigma$} } \cdot {\bf q} \chi \\ \langle B' \vert
J_A^i \vert B \rangle 
&=& N_B N_{B'}^\ast {\chi'}^\dagger \Bigg[ \left( g_1(q^2) +
\frac{M_{B'} - M_B}{M_B + M_{B'}} g_2(q^2) \right) \sigma^i \nonumber
\\ &+& \frac{1}{4 M_B M_{B'}} \bigg(g_3(q^2) - \frac{1}{2} \bigg(
g_1(q^2) \nonumber \\
&+& \frac{M_{B'} - M_B}{M_B + M_{B'}} g_2(q^2) \bigg) \bigg)
q^i \mbox{\boldmath$\sigma$} \cdot {\bf q} \Bigg] \chi
\end{eqnarray}
for the axial-vector current, where $N_B$ ($N_{B'}$) and $\chi$
($\chi'$) are a normalization factor and two-component
non-rela\-ti\-vis\-tic Pauli spinors of the initial (final) baryon state,
respectively. We next introduce a set of auxiliary functions according
to the following definitions
\begin{eqnarray}
\langle B' \vert J_V^0 \vert B \rangle &\equiv& N_B N_{B'}^\ast
{\chi'}^\dagger v_0 \chi \label{eq:Sachs1}\\ \langle B' \vert J_V^i
\vert B \rangle &\equiv& N_B N_{B'}^\ast {\chi'}^\dagger \left( v_V q^i
+ i \epsilon^{ijk} v_A q^j \sigma^k \right) \chi \\ \langle B' \vert
J_A^0 \vert B \rangle &\equiv& N_B N_{B'}^\ast {\chi'}^\dagger a_0
\mbox{\boldmath$\sigma$} \cdot {\bf q} \chi \\ \langle B' \vert J_A^i \vert B
\rangle &\equiv& N_B N_{B'}^\ast {\chi'}^\dagger \left( a_S \sigma^i +
a_T q^i \mbox{\boldmath$\sigma$} \cdot {\bf q} \right) \chi.
\label{eq:Sachs4}
\end{eqnarray}
The functions $v_0$, $v_V$, $v_A$, $a_0$, $a_S$, and $a_T$ are so
called generalized Sachs form factors. The structure of
(\ref{eq:Sachs1}) - (\ref{eq:Sachs4}) can be deduced from
rotational and parity invariance. We also introduce the mass
parameters $\Delta \equiv M_B - M_{B'}$ and $\Sigma \equiv M_B +
M_{B'}$. Identifying the vector functions, we obtain at ${\bf q}^2
\approx 0$
\begin{eqnarray}
v_0 &=& f_1 + \frac{\Delta}{\Sigma} f_3 \label{eq:v0(f)}\\ v_V &=& -
\frac{\Delta}{\Sigma^2 - \Delta^2} \left( f_1 + f_2 \right) +
\frac{1}{\Sigma} f_3 \\ v_A &=& \frac{\Sigma}{\Sigma^2 - \Delta^2}
\left( f_1 + f_2 \right)
\label{eq:vM(f)}
\end{eqnarray}
and solving these equations for $f_i$, where $i=1,2,3$, we get
\begin{eqnarray}
f_1 &=& v_0 - \Delta v_V - \frac{\Delta^2}{\Sigma} v_A
\label{eq:f1(v)}\\
f_2 &=& - v_0 + \Delta v_V + \Sigma v_A
\label{eq:f2(v)}\\
f_3 &=& \Sigma v_V + \Delta v_A, \label{eq:f3(v)}
\end{eqnarray}
at ${\bf q}^2 = 0$, which corresponds to $q^2 = \Delta^2$. Similarly,
for the axial-vector functions, we obtain at ${\bf q}^2 \approx 0$
\begin{eqnarray}
a_0 &=& \frac{\Delta}{\Sigma^2 - \Delta^2} \left( g_3 - g_1 \right) -
\frac{1}{\Sigma} g_2 \label{eq:a0(g)}\\ a_S &=& g_1 -
\frac{\Delta}{\Sigma} g_2 \\ a_T &=& \frac{1}{\Sigma^2 - \Delta^2}
\left( g_3 - \frac{1}{2} \left( g_1 - \frac{\Delta}{\Sigma} g_2
\right)\right)
\label{eq:aT(g)}
\end{eqnarray}
and solving these equations for $g_i$, where $i=1,2,3$, we get
\begin{eqnarray}
g_1 &=& \frac{\Sigma^2 - \Delta^2}{\Sigma^2} \bigg( - \Delta a_0 +
\frac{1}{\Sigma^2 - \Delta^2} \left( \Sigma^2 - \frac{\Delta^2}{2}
\right) a_S \nonumber \\
&+& \Delta^2 a_T \bigg) \label{eq:g1(a)}\\ g_2 &=& -
\frac{1}{\Sigma} \bigg( \left( \Sigma^2 - \Delta^2 \right) a_0 +
\frac{\Delta}{2} a_S \nonumber \\
&-& \left( \Sigma^2 - \Delta^2 \right) \Delta a_T
\bigg) \label{eq:g2(a)}\\ g_3 &=& \frac{1}{2} a_S +
\left( \Sigma^2 - \Delta^2 \right) a_T,
\label{eq:g3(a)}
\end{eqnarray}
at ${\bf q}^2 = 0$, which corresponds to $q^2 = \Delta^2$. It is
important to also keep $f_3$ and $g_3$ non-zero in order to correctly
invert the $v$'s to the $f$'s and the $a$'s to the $g$'s
\cite{cars88}. As mentioned, the $f$'s and $g$'s are true Lorentz
scalar functions, whereas the generalized Sachs form factors are
not. The relations between the $f$'s and $v$'s and the $g$'s and $a$'s
thus depend on the Lorentz-frame in which the calculations are
performed, and therefore, all calculations must be performed in the
same Lorentz-frame. We have made our calculations in the Breit-frame
(which is a good frame \cite{eegh86}), in the non-relativistic limit.

\subsection{The chiral quark model weak form factors}

Next, we calculate the generalized Sachs form factors in the $\chi$QM
to linear order in the symmetry breaking. In the $\chi$QM the form
factors at quark-level are $f_1^q = 1$, $f_2^q = 0$, $f_3^q = 0$,
$g_1^q = g_{a}$, $g_2^q = 0$, and $g_3^q \neq 0$, since, to lowest
order, the $\chi$QM vector current is
\begin{equation}
J_{V,{qq'}}^\mu = \bar{\psi}_{q'} \gamma^\mu \psi_q
\label{eq:VQM}
\end{equation}
and the $\chi$QM axial-vector current is
\begin{equation}
J_{A,{qq'}}^\mu = g_{a}\bar{\psi}_{q'} \gamma^\mu \gamma^5 \psi_q -
f_\Phi \partial^\mu \Phi_{qq'},
\label{eq:AQM2}
\end{equation}
where $g_a$ is the quark axial-vector current coupling constant, and
$\psi_q$, $q = u,d,s$, are Dirac spinors. The parameter $g_{a}$ was
introduced by Manohar and Georgi \cite{mano84} as a possible
``matching parameter'' for the $\chi$QM Lagrangian after spontaneous
symmetry breaking. Later on, we will argue that it should be possible
to put $g_{a}=1$, but for the moment we will keep this parameter free.

The term $- f_\Phi \partial^\mu \Phi_{qq'}$ in the axial-vector
current~(\ref{eq:AQM2}) appears because of the presence of GBs in the
$\chi$QM. Here $f_\Phi$ is the pseudoscalar decay constant and
$\Phi_{qq'}$ is the pseudoscalar field given by
\begin{equation}
\Phi = ({\Phi_{qq'}}) = \left( \begin{array}{ccc}
 	\frac{\pi^0}{\sqrt{2}}+ \frac{\eta}{\sqrt{6}} & \pi^+ & K^+ \\
 	\pi^- & -\frac{\pi^0}{\sqrt{2}}+ \frac{\eta}{\sqrt{6}} & K^0
 	\\ K^- & \bar{K}^0 & - \frac{2\eta}{\sqrt{6}} \end{array}
 	\right). \label{fi}
\end{equation}

The effective Lagrangian for the quark-GB coupling is
\begin{equation}
{\cal L}_{qq'} = i g_8 \bar{\psi}_{q'} \Phi_{q' q} \gamma^{5} \psi_q,
\end{equation}
where $g_8 \equiv g_a (m_q + m_{q'}) / f_\Phi$.

In addition to the octet GBs there is also an SU(3) singlet of $\eta'$
bosons. These are coupled to the quarks with a different strength,
since the theory would otherwise be U(3) symmetric, something that
does not agree with the measurements of the flavor asymmetry measured
by the NMC (New Muon Collaboration) \cite{amau91,arne94} in DIS and
the NA51 Collaboration \cite{bald94} in Drell--Yan production. The
symmetry breaking SU(3) scalar interaction has the form ${\cal
L'}_{qq} = i g_0 \frac{1}{\sqrt{3}} \bar{\psi}_q \eta' \gamma^{5}
\psi_q$, where $g_0$ is the coupling constant for the $\eta'$ bosons.

The effect of this coupling is that the emission of the GBs will in
general flip the spin of the quarks. The interaction of the GBs is
weak enough to be treated by perturbation theory. This means that on
long enough time scales for the low energy parameters to develop we
have
\begin{eqnarray}
u^{\uparrow} & \rightleftharpoons & (d^{\downarrow}+ \pi^+) +
(s^{\downarrow} +K^+) + (u^{\downarrow} + \pi^0, \eta, \eta'),\\
d^{\uparrow}& \rightleftharpoons & (u^{\downarrow}+ \pi^-) +
(s^{\downarrow} +K^0 ) + (d^{\downarrow} + \pi^0 ,\eta ,\eta'),\\
s^{\uparrow} & \rightleftharpoons & (u^{\downarrow}+ K^-) +
(d^{\downarrow} +K^0) + (s^{\downarrow} + \eta ,\eta').
\end{eqnarray}
The probability of transforming a quark with with spin up by one
interaction is given by
\begin{eqnarray}
|\psi(u^\uparrow)|^2 & = &\frac{a}{3}(2+\zeta^2) \hat{u}^\downarrow +a
\hat{d}^\downarrow +a \hat{s}^\downarrow,\\ |\psi(d^\uparrow)|^2 & = &
a \hat{u}^\downarrow + \frac{a}{3}(2+\zeta^2) \hat{d}^\downarrow +a
\hat{s}^\downarrow,\\ |\psi(s^\uparrow)|^2 & = & a \hat{u}^\downarrow
+ a \hat{d}^\downarrow + \frac{a}{3}(2+\zeta^2) \hat{s}^\downarrow,
\end{eqnarray}
where $\zeta \equiv g_0/g_8$ and the coefficient of the
$\hat{q}^\downarrow$, where $q=u,d,s$, should be interpreted as the
probability of creating this quark with spin down by emitting a GB
from a quark with spin up. The parameter $a$ measures the probability
of emission of a GB from a quark. The total probability of GB emission
is $a(8+\zeta^2)/3$.

\begin{figure} 
\epsfig{figure=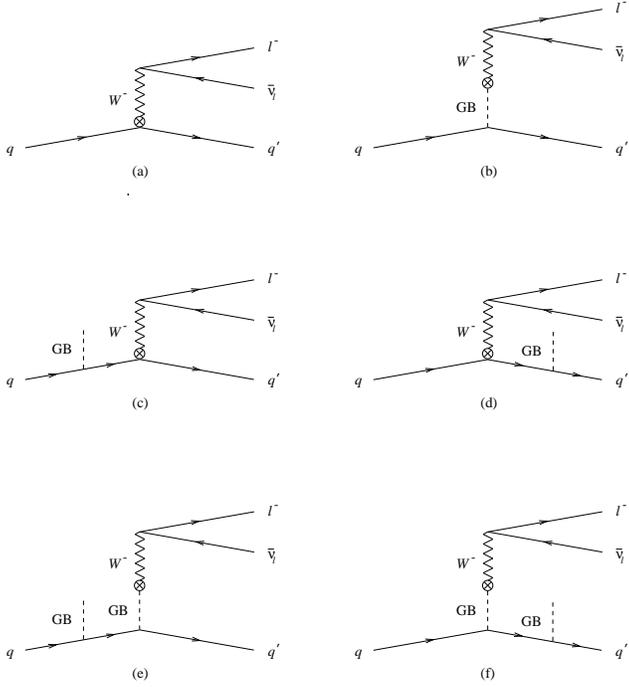,height=9cm}
\caption{Diagrams contributing to the transition
    $q \to q' + l^- + \bar{\nu}_l$. {\bf a}, {\bf b} are zeroth order
    diagrams with respect to $f_\Phi$, {\bf c}--{\bf f} are first order
    diagrams with respect to $f_\Phi$} 
\label{fig:xQM} 
\end{figure}

In Fig.~\ref{fig:xQM}, diagrams (a) and (b) illustrate the two terms
in (\ref{eq:AQM2}). For $\Delta S = 0$ transitions ($d \to u$),
$\Phi_{du} = \pi^-$, and for $\Delta S = 1$ transitions ($s \to u$),
$\Phi_{su} = K^-$. The second term in the axial-vector current will
lead to a non-zero pseudoscalar term (see (\ref{eq:def2}) below),
{\it i.e.} $g_3^q \neq 0$. The diagrams (c) - (f) in the same Figure
illustrate the emission of GBs that can depolarize the quarks and can
even change their flavors.

Spontaneous symmetry breaking in the $\chi$QM will give the mass $m_q$
to the $q$ quark and the mass $m_{\Phi}$ to the pseudoscalar field and
the divergence of the axial-vector current in (\ref{eq:AQM2}) will
be $\partial_{\mu}J_{A,{qq'}}^\mu = f_{\Phi}m^2_{\Phi} \Phi_{qq'}$.
Using the Dirac equation for the quarks on the divergence of the quark
part of the axial-vector current, one obtains
\begin{equation}
\left( \square + {m_\Phi}^2 \right) \Phi_{qq'} = i \frac{m_q +
m_{q'}}{f_\Phi} g_{a}\bar{\psi}_{q'} \gamma^5 \psi_q.
\label{eq:box2}
\end{equation}
The induced pseudoscalar part of the quark axial-vector current matrix
element for the $q \to q' + l^- + \bar{\nu}_l$ decay is defined as
\begin{eqnarray}
\langle q' \vert - f_\Phi \partial^\mu \Phi_{qq'} \vert q \rangle
&\equiv& \bar{u}' \left( \frac{g_3^q}{m_q + m_{q'}} q^\mu \gamma^5
\right) u \nonumber \\
&=& \frac{g_3^q}{m_q + m_{q'}} q^\mu \bar{u}' \gamma^5 u.
\label{eq:def}
\end{eqnarray}
Going over to momentum space, we can solve (\ref{eq:box2}) for
$\Phi_{qq'}$ and insert into $\langle q' \vert - f_\Phi \partial^\mu
\Phi_{qq'}\vert q \rangle$, to obtain
\begin{eqnarray}
\langle q' \vert - f_\Phi \partial^\mu \Phi_{qq'} \vert q \rangle &=&
\frac{m_q+ m_{q'}}{q^2 - m_\Phi^2}g_{a} q^\mu \langle q' \vert
\bar{\psi}_{q'} \gamma^5 \psi_q \vert q \rangle \nonumber \\
&=& \frac{m_q +
m_{q'}}{q^2 - m_\Phi^2} g_{a} q^\mu \bar{u}' \gamma^5 u.
\label{eq:def2}
\end{eqnarray}
This equation corresponds to diagram (b) in Fig.~\ref{fig:xQM}.
Identifying (\ref{eq:def}) and (\ref{eq:def2}), we find that
\begin{equation}
g_3^q = \frac{\sigma^2}{\Delta^2 - m_\Phi^2}g_{a}
\label{eq:g3QM}
\end{equation}
at $q^2 = \Delta^2$. Note that (\ref{eq:g3QM}) is ${\cal O}\left(
f_\Phi^0 \right)$, thus the two diagrams (a) and (b) in
Fig.~\ref{fig:xQM} are of the same order in $f_\Phi$.

We will now make the assumption that the kinetic energy of the
constituent quarks is small enough to allow us to use the static
approximation for them. The advantage of this is that the results will
be less model dependent than by using bound state model wave
functions. The disadvantage is of course that it might be too rough an
approximation. On the other hand, we should understand these
calculations to be done at the same level of approximation for both
the magnetic moments and the weak form factors, since the effective
quark parameters can then be used to relate these observables to each
other. If we change the model for one of these sets of observables,
this would not be possible.

Using the equivalents of (\ref{eq:v0(f)}) - (\ref{eq:vM(f)}) and
(\ref{eq:a0(g)}) - (\ref{eq:aT(g)}), we obtain at quark-level
\begin{eqnarray}
v_0^q &=& 1 \\ v_V^q &=& - \frac{\delta}{\sigma^2 - \delta^2} \\ v_A^q
&=& \frac{\sigma}{\sigma^2 - \delta^2}
\end{eqnarray}
and
\begin{eqnarray}
a_0^q &=& \frac{\delta}{\sigma^2 - \delta^2} \left( g_3^q - g_{a}
\right) \\ a_S^q &=& g_{a} \\ a_T^q &=& \frac{1}{\sigma^2 - \delta^2}
\left( g_3^q - \frac{g_{a}}{2} \right),
\end{eqnarray}
where $\delta \equiv m_q - m_{q'}$, $\sigma \equiv m_q + m_{q'}$, and
$g_3^q$ is the induced pseudoscalar form factor at quark-level.

The quark current operators~(\ref{eq:VQM}) and (\ref{eq:AQM2}) will be
sandwiched between baryon state vectors with (total) spin up in both
the initial and the final states. In the non-relativistic limit, the
current operators then act additively on the three quarks in the
baryons. We will therefore use the Sachs form factors for the quark
currents, and identify the corresponding Sachs form factors for the
baryons by their kinematic structure.

The flavor changing quark transitions can be conveniently expressed by
means of the $\lambda_{qq'}$ matrices, which are combinations of SU(3)
Gell-Mann matrices. For the $\Delta S = 0$ decays ($\lambda_{du}$) and
the $\Delta S = 1$ decays ($\lambda_{su}$), we have
$$
\lambda_{du} = \left( \begin{array}{ccc} 0 & 1 & 0 \\ 0 & 0 & 0 \\ 0 &
0 & 0\end{array} \right) \quad \mbox{\rm and} \quad \lambda_{su} =
\left( \begin{array}{ccc} 0 & 0 & 1 \\ 0 & 0 & 0 \\ 0 & 0 &
0\end{array} \right).
$$

The operators to be sandwiched between the baryonic quark model
states, to obtain the Sachs form factors, are therefore
\begin{eqnarray}
v_{0,qq'} &=& \lambda_{qq'} \otimes 1 \label{eq:v0(f)QM}\\ v_{V,qq'}
&=& - \frac{\delta}{\sigma^2 - \delta^2} \lambda_{qq'} \otimes 1 \\
v_{A,qq'} &=& \frac{\sigma}{\sigma^2 - \delta^2}\lambda_{qq'} \otimes
\sigma^z
\label{eq:vM(f)QM}
\end{eqnarray}
and
\begin{eqnarray}
a_{0,qq'} &=& \frac{\delta}{\sigma^2 - \delta^2} \left(g_3^q - g_{a}
\right) \lambda_{qq'}\otimes \sigma^z
\label{eq:a0(g)QM}\\
a_{S,qq'} &=& g_{a} \lambda_{qq'}\otimes \sigma^z \\ a_{T,qq'} &=&
\frac{1}{\sigma^2 - \delta^2} \left(g_3^q - \frac{g_{a}}{2}\right)
\lambda_{qq'}\otimes \sigma^z.
\label{eq:aT(g)QM}
\end{eqnarray}
The $\lambda_{qq'}$ matrix effectuates the flavor transition and the
$\sigma^z$ operator measures the spin polarizations of the quarks in
the baryons.

In a given type of transition, say $\Delta S =1$, the active quark
masses are the same and the spectator quark masses do not enter
explicitly in the calculations. Introducing the notation $f_1^{\rm QM}
\equiv \langle B' \vert \lambda_{qq'} \otimes 1 \vert B \rangle$ and
$g_1^{\rm QM} \equiv \langle B' \vert \lambda_{qq'} \otimes \sigma^z
\vert B \rangle$, we can identify the $v$'s in the baryonic matrix
element and insert them into (\ref{eq:f1(v)}) - (\ref{eq:f3(v)}),
to obtain
\begin{equation}
f_1 = \left( 1 + \frac{\Delta}{\sigma^2 - \delta^2} \left( \delta -
\frac{\sigma}{\Sigma} \Delta \frac{g_1^{\rm QM}}{f_1^{\rm QM}} \right)
\right) f_1^{\rm QM},
\label{eq:f_1}
\end{equation}
\begin{equation}
f_2 = \left( \frac{1}{\sigma^2 - \delta^2} \left( \Sigma \sigma
\frac{g_1^{\rm QM}}{f_1^{\rm QM}} - \Delta \delta \right) - 1 \right)
f_1^{\rm QM},
\label{eq:f_2}
\end{equation}
and
\begin{equation}
f_3 = \frac{1}{\sigma^2 - \delta^2} \left( \Delta \sigma
\frac{g_1^{\rm QM}}{f_1^{\rm QM}} - \Sigma \delta \right) f_1^{\rm
QM}.
\label{eq:f_3}
\end{equation}
In a similar way, we can identify the $a$'s and insert them into
(\ref{eq:g1(a)}) - (\ref{eq:g3(a)}). The result is
\begin{eqnarray}
g_1 &=& \bigg( 1 + \frac{\Sigma^2 - \Delta^2}{\Sigma^2}\bigg(
\frac{\Delta \delta}{\sigma^2 - \delta^2} - \frac{\Delta^2}{2} \bigg(
\frac{1}{\Sigma^2 - \Delta^2} \nonumber \\
&+& \frac{1}{\sigma^2 - \delta^2} \bigg)
+ \frac{\Delta}{\sigma^2 - \delta^2} \left( \Delta - \delta \right)
\frac{g_3^q}{g_{a}} \bigg) \bigg) g_{a} g_1^{\rm QM},
\label{eq:g_1}
\end{eqnarray}
\begin{eqnarray}
g_2 &=& \frac{1}{\Sigma} \bigg( \frac{\Sigma^2 - \Delta^2}{\sigma^2 -
\delta^2} \left( \delta - \frac{\Delta}{2} \right) - \frac{\Delta}{2}
\nonumber \\
&+& \frac{\Sigma^2 - \Delta^2}{\sigma^2 - \delta^2} \left( \Delta -
\delta \right) \frac{g_3^q}{g_{a}} \bigg) g_{a} g_1^{\rm QM},
\label{eq:g_2}
\end{eqnarray}
and
\begin{equation}
g_3 = \left( \frac{1}{2} \left( 1 - \frac{\Sigma^2 -
\Delta^2}{\sigma^2 - \delta^2} \right) + \frac{\Sigma^2 -
\Delta^2}{\sigma^2 - \delta^2} \frac{g_3^q}{g_{a}} \right) g_{a}
g_1^{\rm QM}.
\label{eq:g_3}
\end{equation}
The weak currents on baryon-level and quark-level have to be
calculated in the same reference frame in order to maintain Lorentz
invariance of the weak form factors $f_i$ and $g_i$, where $i=1,2,3$
\cite{eegh86}.

The final result will contain a multiplicative factor from the wave
function overlap, contributing to the so called wave function
mismatch. Actually, this mismatch comes about from two different
sources.

The first one is the recoil effect, that for non-relativistic systems
is proportional to the matrix element of the spherical Bessel function
$j_{0}(\Delta r)$, where $r$ is the radial coordinate. If we consider
an expansion in $\delta$ and $\Delta$, we get
$$
j_{0}(\Delta r) \equiv \frac{\sin (\Delta r)}{\Delta r} = 1 - \frac{1}{6}
\Delta^2 r^2 + \cdots.
$$
The contribution from this term that is
different from $1$ is therefore ${\cal O}(\Delta^2)$. For spherically
symmetric wave functions ($S$-waves), the lowest order relativistic
effects in the kinematic terms can also be shown to be ${\cal
O}(\delta^2)$.

Secondly, we have the contribution from the overlap between two wave
functions that have different quark \\ masses. By expanding the wave
function in the quark mass difference $\delta$, it can easily be shown
that the deviation of this effect from $1$ is also ${\cal
O}(\delta^2)$. Since we are calculating only the linear part of the
symmetry breaking in the weak form factors, we will therefore in the
following neglect the wave function mismatch.

Define now the parameters $E \equiv \Delta/\Sigma$ and $\epsilon
\equiv \delta/\sigma $. If we express (\ref{eq:f_1}) -
(\ref{eq:f_3}) and (\ref{eq:g_1}) - (\ref{eq:g_3}) in $\Sigma$, $E$,
$\sigma$, and $\epsilon$ and neglect all terms which are proportional
to $E^2$, $\epsilon^2$, and $E \epsilon$, we obtain
\begin{eqnarray}
f_1 &=& f_1^{\rm QM} \label{eq:f1app}\\ f_2 &=& \left(
\frac{\Sigma}{\sigma} G_A - 1 \right) f_1^{\rm QM}
\label{eq:f2app}\\
f_3 &=& \frac{\Sigma}{\sigma} \left( E G_A - \epsilon \right) f_1^{\rm
QM}
\label{eq:f3app}
\end{eqnarray}
and
\begin{eqnarray}
g_1 &=& g_{a} g_1^{\rm QM} \label{eq:g1app}\\ g_2 &=& \left(
\frac{\Sigma}{\sigma} \epsilon - \frac{1}{2} \left(1 +
\frac{\Sigma^2}{\sigma^2} \right) E \right) g_{a} g_1^{\rm QM}
\label{eq:g2app}\\ g_3 &=& \left( \frac{1}{2} \left( 1 -
\frac{\Sigma^2}{\sigma^2} \right) + \frac{\Sigma^2}{\sigma^2}
\frac{g_3^q}{g_{a}} \right) g_{a} g_1^{\rm QM},
\label{eq:g3app}
\end{eqnarray}
where $G_A \equiv g_1^{\rm QM}/f_1^{\rm QM}$. In this result we have
also deleted the term in $g_{2}$ proportional to $g_{3}^{q}$, since it
should be absent on physical grounds. The current piece containing the
$g_{2}$ form factor is orthogonal to $q_{\mu}$, whereas $g_{3}^{q}$ is
proportional to the divergence of the axial-vector current.

We note that the first-class current form factors $f_1$, $f_2$, $g_1$,
and $g_3$ only contain terms with even powers of $E$ and $\epsilon$,
while the second-class current form factors $f_3$ and $g_2$ only
contain terms with odd powers of $E$ and $\epsilon$. This follows from
the Ademollo--Gatto theorem \cite{adem64,eegh86}. The above
expressions for $f_i$ and $g_i$, where $i=1,2,3$, in
(\ref{eq:f1app}) - (\ref{eq:g3app}) are evaluated at $q^2 =
\Delta^2$.

Using (\ref{eq:g3QM}), this means that (\ref{eq:g1app}) -
(\ref{eq:g3app}) now can be expressed as
\begin{eqnarray}
g_1 &=& g_{a}g_1^{\rm QM} \label{eq:g1appchi}\\ g_2 &=& \left(
\frac{\Sigma}{\sigma} \epsilon - \frac{1}{2} \left(1 +
\frac{\Sigma^2}{\sigma^2} \right) E \right) g_{a} g_1^{\rm QM}
\label{eq:g2appchi}\\ g_3 &=& \left( \frac{1}{2} \left( 1 -
\frac{\Sigma^2}{\sigma^2} \right) + \frac{\Sigma^2}{\Delta^2 -
m_\Phi^2} \right)g_{a} g_1^{\rm QM},
\label{eq:g3appchi}
\end{eqnarray}
in the $\chi$QM. We will keep the $\Delta^2$ term in the denominator
in (\ref{eq:g3appchi}), since there is no natural expansion
parameter in this case.

The usual way to obtain the values of the form factors $f_1$ and $g_1$
at $q^2 = 0$ is to use the empirical dipole forms for the $q^2$
dependence of these form factors. It is easy to see that the
difference between, say, $f_{1}(0)$ and $f_{1}(\Delta^2)$ is ${\cal
O}(\Delta^2)$. But, since the form factors $f_1$ and $g_1$ are only
valid up to quadratic terms in the mass differences, we will neglect
the $q^2$ dependence from the empirical dipole forms and put $f_1 =
f_1(\Delta^2) \approx f_1(0)$ and $g_1 = g_1(\Delta^2) \approx
g_1(0)$.

The $q^2$ dependences of the other form factors $f_2$, $f_3$, $g_2$,
and $g_3$ are also neglected, since these dependences cannot be
decided with the current level of experimental precision.

In the $\chi$QM, the effective quark masses can be determined from the
fitted value of $\mu_d$, which in the $\chi$QM is $\mu_d \approx -1.35
\, \mu_N$ \cite{lind97}. Using this value together with the formulas
from the magnetic moments, $\mu_u = -2 \mu_d$, and $\mu_s = 2\mu_d/3$
\cite{lind97}, the effective quark masses in the $\chi$QM are
$m_u^{\rm eff} = m_d^{\rm eff} = m^{\rm eff} \approx 230 \, {\rm MeV}$
and $m_s^{\rm eff} = 3 m^{\rm eff}/ 2\approx 350 \, {\rm MeV}$. For
the form factor $g_{3}$ we have used $m_{\pi}=140$ MeV and $m_{K}=490$
MeV. This seems to be consistent with the pole in $g_{3}$ coming from
$g_{3}^{q}$ being identified with the pion (kaon) pole in $g_{3}$ from
dispersion relations.

\subsection{The weak axial-vector form factors}
\label{sec:GA}

The weak axial-vector form factors $G_A = g_{1}^{\rm QM}/f_{1}^{\rm
QM}$ can be obtained from the SU(6) quark model expressions for
$f_{1}$ and $g_{1}$ expressed in terms of the parameters $F$ and $D$
\cite{rent90}. In the $\chi$QM, the weak axial-vector form factors
$G_A$ are expressed in the quark spin polarizations of the proton,
{\it i.e.} $\Delta u$, $\Delta d$, and $\Delta s$. These spin
polarizations differ considerably from the ones in the SU(6) quark
model due to the depolarization of the quark spins by the GBs. The
spin polarizations in the $\chi$QM are calculated to ${\cal
O}(f_\Phi)$, {\it i.e.} with one GB emission. They are \cite{chen95}
\begin{eqnarray}
\Delta u &=& \frac{4}{3} - \frac{a}{9} \left( 8 \zeta^2 + 37
\right),\\ \Delta d &=& - \frac{1}{3} + \frac{2 a}{9} \left( \zeta^2 -
1 \right),\\ \Delta s &=& - a.
\end{eqnarray}
For values of $\Delta u$, $\Delta d$, and $\Delta s$ in the $\chi$QM,
see Table~\ref{tab:pol}. Using the relations $F = \frac{1}{2} (\Delta
u - \Delta s)$ and $D = \frac{1}{2} (\Delta u - 2 \Delta d + \Delta
s)$ \cite{clos93}, we have
\begin{eqnarray}
G_A^{np} &=& \Delta u - \Delta d\\ G_A^{\Sigma^- \Sigma^0} &=&
\frac{1}{2} (\Delta u -\Delta s)\\ {g_1^{\rm QM}}^{\Sigma^\pm \Lambda}
&=& \frac{1}{\sqrt{6}} (\Delta u - 2 \Delta d + \Delta s)\\ G_A^{\Xi^-
\Xi^0} &=& \Delta d - \Delta s
\end{eqnarray}
for the $\Delta S = 0$ decays and
\begin{eqnarray}
G_A^{\Sigma^- n} &=& \Delta d - \Delta s\\ G_A^{\Xi^- \Sigma^0} &=&
\Delta u - \Delta d\\ G_A^{\Xi^- \Lambda} &=& \frac{1}{3} (\Delta u +
\Delta d - 2 \Delta s)\\ G_A^{\Lambda p} &=& \frac{1}{3} (2 \Delta u -
\Delta d - \Delta s)\\ G_A^{\Xi^0 \Sigma^+} &=& \Delta u - \Delta d
\end{eqnarray}
for the $\Delta S = 1$ decays.

The $\Sigma^0 \to \Sigma^+ + l^- + \bar{\nu}_l$ and $\Sigma^0 \to p +
l^- + \bar{\nu}_l$ decays cannot be observed, since the
electromagnetic decay $\Sigma^0 \to \Lambda + \gamma$ is predominant.
The corresponding $G_A$'s are therefore not listed above.

The values of the $G_A^{BB'}$'s for the $\chi$QM are listed in
Table~\ref{tab:GA}, where for reference also the axial-vector form
factors of the NQM are displayed.

The weak axial-vector form factor $g_A$ is defined as
\begin{equation}
g_A \equiv \frac{g_1}{f_1}.
\end{equation}
For the weak vector form factor $f_{1}$, the $\chi$QM gives the same
result as the ordinary NQM. The appropriate values can be found in
Table~\ref{tab:fg}.
We thus obtain the simple result
\begin{equation}
g_A = g_{a}G_A.
\end{equation}

It has been argued by Weinberg \cite{wein90}, that not only in QCD, but
also in the effective Lagrangians, one should expect $g_{a}=1$, since
the matrix element algebra of the axial-vector currents between color
quark states should be saturated by the single quark state to leading
order in $1/N_{c}$, where $N_c$ is the number of colors. This leads to
$g_{a}=1$. Compare with the Adler--Weisberger relation, that relates
the deviation of $g_{A}^{np}$ from 1 to the presence of excited
intermediate states, like the $\Delta(1232)$ resonance, in the
saturation of the sum-rule. The subleading order corrections that come
from quark-GB interactions \cite{wein91} are taken care of by the
depolarization of the quark spins due to GB emission above. The
renormalization of the axial-vector form factor for $g_{A}^{np}$ from
its SU(6) value of $5/3$ to its experimental value of $1.26$ should
then come entirely from the change in spin polarization due to the
GBs, otherwise there is a risk for double counting. This attitude for
$g_{a}$ in the $\chi$QM has also been taken by other authors
\cite{chen95,lind97,chen97,webe97,webe972} and will be adopted
here. See, however, also Ref.~\cite{peri93}.

Expressed in terms of matrix elements, the weak axial-vector form
factors $g_A^{BB'}$ in the $\chi$QM will then equal to $g_A^{BB'} =
G_A^{BB'}(\Delta u,\Delta d,\Delta s)$ as given above.

\subsection{The ratio $\mbox{\boldmath$\rho_f$}$ and the ``weak magnetism''}
\label{sub:ratio}

We next turn to the ``weak magnetism'' form factor $\rho_f$, which is
defined as
\begin{equation}
\rho_f \equiv \frac{f_2}{f_1}.
\label{eq:rhof}
\end{equation}
Inserting (\ref{eq:f1app}) and (\ref{eq:f2app}) in
(\ref{eq:rhof}), we obtain
\begin{equation}
\rho_f = \frac{\Sigma}{\sigma} G_A - 1.
\end{equation}
The formula above can be transformed into an expression in terms of
the magnetic moments of the baryons. For example, for the $n \to p +
l^- + \bar{\nu}_l$ decay, we can show, using $\mu_p = \Delta u \,
\mu_u + \Delta d \, \mu_d +\Delta s \, \mu_s$ and the corresponding
formula for $\mu_n$, that
\begin{eqnarray}
\rho_f^{n p} &=& \frac{1}{2} \left( 1 + \frac{M_n}{M_p} \right) \left(
\mu_p - \mu_n \right) \frac{1}{\mu_N} - 1 \nonumber \\
&\approx& \left( \mu_p - \mu_n
\right) \frac{1}{\mu_N} - 1.
\label{eq:CVC}
\end{eqnarray}
Here we have used the expression $G_A^{n p} = \Delta u - \Delta d$
from Subsection~\ref{sec:GA} above and $\mu_u = - 2
\mu_d$. Equation~(\ref{eq:CVC}) is exactly the conserved vector
current (CVC) formula for the $n \to p + l^- + \bar{\nu}_l$
decay. Using the $\chi$QM values $\mu_p \approx 2.67 \, \mu_N$ and
$\mu_n \approx -1.86 \, \mu_N$ \cite{lind97}, we thus obtain
$\rho_f^{n p} \approx 3.53$, in agreement with the direct calculation
(see Table~\ref{tab:rho}).

The expression for $\rho_{f}$ above is closely related to the
corresponding formula for the magnetic moments $\mu_B$ of the octet
baryons used in earlier studies. In the same approximation as here, we
have
\begin{equation}
	 f_{2}(0)/f_{1}(0) = \Sigma \mu_{B} -1 =\Sigma \sum_{q=u,d,s}
	\frac{e_q}{2m_{q}} \Delta q -1, \label{mag}
\end{equation}
where $e_{q}$ is the quark charge.

When these expressions are fitted to the baryon magnetic moments, the
quark masses appear as effective mas\-ses, and the parametric dependence
of the quark spin polarization $\Delta q$ on the emission probability
$a$ of GBs incorporates effects of relativistic corrections and other
possible dynamical effects on the magnetic moments \cite{dann97}. When
these effects are taken into account directly, in terms of a changed
structure of the currents, the fits become quite bad
\cite{webe972}. At the present time the above treatment is therefore
probably the best one can hope for.

\subsection{The ratio $\mbox{\boldmath$\rho_g$}$ and the weak form
factor $\mbox{\boldmath$g_{PT}$}$}
\label{sub:ratioandfor}

The ratio $\rho_g$ is defined as
\begin{equation}
\rho_g \equiv\frac{g_2}{g_1}.
\end{equation}
Thus $\rho_g$ is obtained by dividing (\ref{eq:g2appchi}) by
(\ref{eq:g1appchi})
\begin{equation}
\rho_g = \frac{\Sigma}{\sigma} \epsilon - \frac{1}{2} \left(1 +
\frac{\Sigma^2}{\sigma^2} \right) E.
\label{eq:rhogchi}
\end{equation}
The ratio $\rho_g$ depends only on the masses of the quarks $q$, $q'$
and the baryons $B$, $B'$, and not on $g_{a}$.

The weak induced pseudotensor form factor $g_{PT}$ is defined as
\begin{equation}
g_{PT} \equiv \frac{g_2}{f_1}.
\end{equation}
We then have
\begin{equation}
g_{PT} = \left( \frac{\Sigma}{\sigma} \epsilon - \frac{1}{2} \left(1 +
\frac{\Sigma^2}{\sigma^2} \right) E \right) G_A.
\end{equation}
Using the ratio $\rho_g \equiv g_{2}/g_{1}$, we can relate the form
factor $g_{PT}$ to the form factor $g_A$ according to
\begin{equation}
g_{PT} = \frac{g_2}{f_1} = \frac{g_2}{g_1} \frac{g_1}{f_1} = \rho_g
g_A.
\end{equation}
The matrix elements of the weak induced pseudotensor form factor
$g_{PT}$ are thus given by
\begin{equation}
g_{PT}^{BB'} = \rho_g^{BB'} G_A^{BB'}.
\end{equation}

Since different signs for $\rho_{g}$ are obtained in different models
(see Table~\ref{tab:eta}) we would like to see if we can understand
this feature from our estimate. Inspection of (\ref{eq:rhogchi})
shows that its sign will depend upon a balance between the term
proportional to $\epsilon$ and the one proportional to $E$.

For the $\Delta S = 0$ $\Sigma^{\pm} \to \Lambda$ transitions
$\epsilon =0$ so $\rho_{g}$ is negative. This is consistent with the
values presented by all authors and affirms that the same sign
convention is used.

For the $\Delta S = 1$ transitions $\epsilon \neq 0$ and the situation
depends on the balance between the terms. Since $(1 +
\Sigma^2/\sigma^2 ) \approx \Sigma^2/\sigma^2$ for these decays, the
sign of $\rho_{g}$ depends on the sign of $\delta - \Delta /2$. This
value depends evidently upon the models used. In our case the sign is
negative for the $\Sigma^{-} \to n$ transition and positive for the
others.

A similar remark applies to the form factor ratio $f_{3}/f_{1}$. Its
sign is also dependent upon a balance between two terms. For the
$\Delta S = 0$ $\Sigma^{\pm} \rightarrow \Lambda$ transitions, where
$\epsilon=0$, we have $f_{3} = g^{{\rm QM}}_{1}\Delta/\sigma$, which
is positive. For the $\Delta S =1$ transitions we can only say for
sure that it must be negative for decays with negative $G_{A}$. Since
it is not possible at present to measure $f_{3}$ we will not study it
any further.

Also the form factor $g_{3}$ is not possible to measure at present,
although the pole term makes it quite large.

In our calculations we have $\delta \approx 120$ MeV and $\epsilon
\approx 0.20$ for the $\Delta S = 1$ transitions. This means that some
of the form factors should be considered as estimates rather than
calculations. Nevertheless, such estimates are often much better than
one would expect. In particular, as has been mentioned above, the
ratios $\Delta q/m_{q}$, where $\Delta q$ is the spin polarization and
$m_{q}$ the effective mass of a quark with flavor $q$, are well
determined from the magnetic moment calculations, and should reproduce
the different weak form factors well. In our opinion, the over all
performance of the $\chi$QM is quite good, and with one possible
exception, it reproduces the experimental data.

\section{Numerical results}
\label{sec:numres}

\subsection{Experimental values of the weak axial-vector form factors}

The measured weak axial-vector form factor, $g_A^{\rm exp}$, is often
a superposition of the theoretical weak axial-vector form factor $g_A$
and the theoretical weak induced pseudotensor form factor $g_{PT}$,
since one assumes that the form factor $g_{2}$ is zero in the analysis
of data. Thus from the Gordon equality~(\ref{eq:gordon}), one gets
\begin{equation}
g_A^{\rm exp} = g_A - E g_{PT},
\label{eq:gAexpf}
\end{equation}
where $E$ is given by
\begin{equation}
E \equiv \frac{M_{B} - M_{B'}}{M_B + M_{B'}}.
\label{eq:A}
\end{equation}
As a quasi-experimental value for $g_A$ one could take the value
obtained by solving (\ref{eq:gAexpf}) for $g_A$ and inserting our
theoretical prediction for $g_{PT}$. Thus,
\begin{equation}
g_A^{\rm quasi} \equiv g_A^{\rm exp} + E g_{PT}^{{\rm theory}}.
\label{eq:gAexp_cit}
\end{equation}
However, since $E$ is quite small, $E\leq 0.12$, and $g_{PT}$ is also
small, the term $Eg_{PT}$ is negligible in our approximation. This is
consistent with $Eg_{PT}$ being ${\cal O}(E^{2}, E\epsilon)$. The
experimental values of the weak axial-vector form factors
${g_A^{BB'}}^{\rm exp}$ are presented in Table~\ref{tab:gagpt}.

For the $\Sigma^- \to n + l^- + \bar{\nu}_l$ decay, we have
$E^{\Sigma^- n} \approx 0.12$ according to (\ref{eq:A}). Hsueh
{\it et al.} \cite{hsue88} have measured ${g_A^{\Sigma^- n}}^{\rm exp}
= -0.327 \pm 0.007 \pm 0.019$ in a single parameter fit, which
corresponds to (\ref{eq:gAexpf}), and also independently
$g_A^{\Sigma^- n} = -0.20 \pm 0.08$ and ${g_{PT}^{\Sigma^- n}}^{\rm
Hsueh} = 0.56 \pm 0.37$. Hsueh {\it et al.} use a definition of
$g_{PT}$ different from ours, and the definitions are related to each
other by the formula
\begin{equation}
g_{PT}^{\Sigma^- n} = \left( 1 + \frac{M_n}{M_{\Sigma^-}} \right)
{g_{PT}^{\Sigma^- n}}^{\rm Hsueh}.
\label{eq:gPTsmn}
\end{equation}
Equation~(\ref{eq:gPTsmn}) gives $g_{PT}^{\Sigma^- n} = 1.00 \pm
0.66$. Using the definition of $\rho_g$, we now get $\rho_g^{\Sigma^-
n} = g_{PT}^{\Sigma^- n}/g_A^{\Sigma^- n} = -5.0 \pm 3.9$. None of
the presented models in the tables are able to reproduce this value.

\subsection{Discussion}

In Table~\ref{tab:gagpt} we present the $\chi$QM values for the
$g_A$'s. The value of $g_A^{np}$ is slightly low in the $\chi$QM. This
indicates that the theoretical values are still maybe only within
about $10 \%$ of the experimental ones. It is also possible that a
fine tuning of the value for the parameter that measures the strength
of the GB emission could bring the value up. Nevertheless, the
agreement between the experimental values and the model is quite
encouraging and represents a clear improvement over the NQM values.

In the SU(6) model, the value $g_{A}^{np}=5/3$ is related to the value
of $g_{A}$ for the transition $p \to \Delta^{++}$, when the
axial-vector matrix element algebra is saturated with the octet and
decuplet \cite{gers66}.

The improvement of $g_{A}^{np}$ in the $\chi$QM is due to the effect
of the GB emission from the quarks before or after the weak
interaction. This changes the matrix element algebra of the
axial-vector currents that fixes the value of $g_{A}$, since both
before and after the interaction the quark amplitude in the baryonic
states are not in pure SU(6) representations, but rather in a mixture
of such states, not only of different spins, but also of different
flavors.

When it comes to the other form factors the situation is as follows.

For the $\rho_f$ ratios there are more experimental data than for the
$\rho_g$ ratios. Let us therefore consider Table~\ref{tab:rho}. All
values obtained for the $\rho_f$'s in the $\chi$QM lie within the
experimental errors, where experimental data exist. (The experimental
results have large errors, though.) The CVC values listed are in a
way half experimental results, since they use the measured values of
the anomalous magnetic moments for the nucleons as input data to
calculate these values. All calculated values for the $\chi$QM have
the same sign as the CVC values and they are also close in
magnitude. This is of course related to the fact that the form factors
are calculated in the same approximation as the magnetic moments in
earlier studies, and the parameters from these calculations are used
here. For some cases we can see that $\rho_{f}(\chi{\rm QM}) \approx
\rho_{f}({\rm CVC})$, as for the neutron decay. For other decays the
$\rho_{f}$'s of the $\chi$QM incorporate effects of vector current
non-conservation due to the mass differences between the
isomultiplets.

Let us then consider Table~\ref{tab:eta}. Unfortunately, only one of
the $\rho_g$'s, namely $\rho_g^{\Sigma^- n}$, has been measured
experimentally. As mentioned before, this was done by Hsueh {\it et
al.} They found $\rho_g^{\Sigma^- n} = -5.0 \pm 3.9$ (in our
conventions). Theoretically, our estimate gives the value $-0.143$ in
the $\chi$QM, and this is not in agreement with the experimental
value. However, also the values of all other models are outside the
experimental range. Taken at face value, the result for
$\rho_g^{\Sigma^- n}$ as measured by Hsueh {\it et al.} \cite{hsue88}
would tend to favor models with negative values for the
$\rho_{g}$. However, one should perhaps await further measurements
before taking a stand, since the error is quite large, and one more
standard deviation would allow for models with positive $\rho_{g}$.

For the other $\rho_g$'s with $\Delta S =1$, we can only compare our
predictions with previous model calculations. We get a positive sign
for these $\rho_g$'s in agreement with the MIT and $\chi{\rm
QSM_{br}}$ models. The other models have negative signs for the
$\rho_g$'s.

Finally, we present in Table~\ref{tab:gagpt} model estimates for the
$g_{PT}$'s calculated using the values of $\rho_{g}$ and $g_{A}$
presented above. The $\chi$QM value for $g_{PT}^{\Sigma^- n}$, which
is the only measured form factor, is too small compared to the
experiment.

The over all picture of our theoretical estimates for the $\chi$QM
are, apart from the measured value of the form factor $g_{PT}$ and the
value of $\rho_{g}$ for the $\Sigma^{-} \to n$ transition, in good
agreement with the existing experimental data.

\section{Summary and conclusions}
\label{sec:SC}
We have presented a study of the baryonic weak vector and axial-vector
form factors in the spirit of the chiral quark model. The results are
presented in Tables~\ref{tab:fg} - \ref{tab:gagpt}, and the over all
agreement with existing data is satisfactory and represents a clear
improvement with respect to the non-relativistic quark model.

The experimental axial-vector form factors, corrected for the possible
non-zero values of $g_{PT}$, are of importance in {\it e.g.} the
analysis of the quark spin polarizations of the nucleon. Our study
supports the assumption that these form factors are small. The
second-class form factors $f_{3}$ and $g_{2}$ are also highly model
dependent.

The present investigation has used the SU(3) symmetric coupling in the
chiral quark model and the static approximation for the quarks as a
first approximation. A natural improvement would be to incorporate
lowest order non-static effects and further SU(3) symmetry breaking
effects \cite{chen97,song97}, to obtain better agreement with
experimental data. In particular, we expect that this would lead to a
closer agreement with the $\rho_f$ ratios from the conserved vector
current theory, since symmetry breaking can better account for the
octet baryon magnetic moments \cite{lind97}. SU(3) symmetry breaking
also leads to better agreement for $g_A^{n p}$
\cite{lind97,chen97,song97}.

Finally, we think that it would be quite interesting to have more
measurements of $\rho_g$ for various transitions, since this parameter
might help to distinguish between different models.

\begin{acknowledgement}
{\it Acknowledgements.} We would like to thank Johan Linde for useful
comments on the 
manuscript. This work was supported by the Swedish Natural Science
Research Council (NFR), Contract No. F-AA/FU03281-310. Support for
this work was also provided by the Engineer Ernst Johnson Foundation (T.O.).
\end{acknowledgement}

\newpage

\begin{threeparttable}
\caption{Quark spin polarizations. $\Delta \Sigma$ is the total quark
spin polarization, {\it i.e.} $\Delta \Sigma = \Delta u + \Delta d +
\Delta s$. The experimental values have been obtained from
\protect\cite{elli95}.
The data for the NQM and the $\chi$QM can been obtained from
\protect\cite{chen95,lind97}}
\begin{tabular}{lrrr}
\hline
\\
Quantity & Experimental value & NQM & $\chi$QM\\
\hline
$\Delta u$ & $0.83 \pm 0.03$ \tnote{a} & $\frac{4}{3}$ & $0.79$\\
$\Delta d$ & $-0.43 \pm 0.03$ \tnote{a} & $-\frac{1}{3}$ & $-0.32$\\
$\Delta s$ & $-0.10 \pm 0.03$ \tnote{a} & $0$ & $-0.10$\\
$\Delta \Sigma$ & $0.31 \pm 0.07$ \tnote{a} & $1$ & $0.37$\\
\hline
\end{tabular}
\begin{tablenotes}
\item[a]{\footnotesize Obtained with $g_A^{np} \approx 1.26$ and $\Delta u +
\Delta d - 2 \Delta s \approx 0.60$}
\end{tablenotes}
\label{tab:pol}
\end{threeparttable}

\begin{threeparttable}
\caption{Weak axial-vector form factors, $G_A^{BB'}$. The values in
the NQM column are the SU(6) values for the weak axial-vector form
factors and the values in the $\chi$QM column are obtained from the
quark spin polarizations given in Table~\ref{tab:pol}.
${g_1^{\rm QM}}^{\Sigma^\pm \Lambda}$ are given instead of $G_A^{\Sigma^\pm
\Lambda}$, since ${f_1^{\rm QM}}^{\Sigma^\pm \Lambda} = 0$}
\begin{tabular}{lrr}
\hline
\\
Quantity & NQM & $\chi$QM\\
\hline
$G_A^{np}$ & $\frac{5}{3}$ & $1.12$\\
$G_A^{\Sigma^- \Sigma^0}$ & $\frac{2}{3}$ & $0.45$\\
${g_1^{\rm QM}}^{\Sigma^\pm \Lambda}$ & $\sqrt{\frac{2}{3}}$ & $0.55$\\
$G_A^{\Xi^- \Xi^0}$& $-\frac{1}{3}$ & $-0.22$\\
\hline
$G_A^{\Sigma^- n}$ & $-\frac{1}{3}$ & $-0.22$\\
$G_A^{\Xi^- \Sigma^0}$ & $\frac{5}{3}$ & $1.12$\\
$G_A^{\Xi^- \Lambda}$ & $\frac{1}{3}$ & $0.22$\\
$G_A^{\Lambda p}$ & $1$ & $0.67$\\
$G_A^{\Xi^0 \Sigma^+}$ & $\frac{5}{3}$ & $1.12$\\
\hline
\end{tabular}
\label{tab:GA}
\end{threeparttable}

\begin{table*}
\begin{threeparttable}
\caption{The weak form factors $f_i$ and $g_i$, where $i=1,2,3$ in the
$\chi$QM.}
\begin{tabular}{lcccccc}
\hline
\\
Decay & $f_1$ & $f_2$ & $f_3$ & $g_1$ & $g_2$ & $g_3$\\
\hline
$n \to p$ & $1.00$ & $3.52$ & $0$ \tnote{a} & $1.12$ & $0$
\tnote{a} & $-210$\\
$\Sigma^- \to \Sigma^0$ & $1.41$ & $1.84$ & $0$ \tnote{a} & $0.63$ &
$0$ \tnote{a} & $-190$\\
$\Sigma^- \to \Lambda$ & $0$ & $2.73$ & $0.10$ & $0.55$ & $-0.25$ &
$-240$\\
$\Sigma^+ \to \Lambda$ & $0$ & $2.72$ & $0.09$ & $0.55$ & $-0.22$ &
$-210$\\
$\Xi^- \to \Xi^0$ & $1.00$ & $-2.27$ & $0$ \tnote{a} &
$-0.22$ & $0$ \tnote{a} & $83$\\
\hline
$\Sigma^- \to n$ & $-1.00$ & $1.82$ & $0.84$ & $0.22$ & $-0.03$ &
$-7.2$\\
$\Xi^- \to \Sigma^0$ & $0.71$ & $2.72$ & $-0.44$ & $0.79$ & $0.28$
& $-29$\\
$\Xi^- \to \Lambda$ & $1.22$ & $-0.07$ & $-0.93$ & $0.27$ &
$0.01$ & $-10$\\
$\Lambda \to p$ & $-1.22$ & $-1.68$ & $0.62$ & $-0.82$ & $-0.10$ & $21$\\
$\Xi^0 \to\Sigma^+$ & $1.00$ & $3.83$ & $-0.62$ & $1.12$ & $0.42$ &
$-41$\\
\hline
\end{tabular}
\begin{tablenotes}
\item[a]{The mass difference for baryons in the same isospin multiplet
has been neglected}
\end{tablenotes}
\label{tab:fg}
\end{threeparttable}
\end{table*}

\begin{table*}
\begin{threeparttable}
\caption{The ratios $\rho_f^{BB'} \equiv
\frac{f_2^{BB'}}{f_1^{BB'}}$. The experimental values have been
obtained from
\protect\cite{bour82} (see also \protect\cite{gail84}).
The CVC results use as input the
experimental values of the anomalous magnetic moments $\mu_p^a \equiv
\mu_p - 1 \approx 1.793 \mu_N$ and $\mu_n^a \equiv \mu_n \approx
-1.913 \mu_N$, where $\mu_N$ is the nuclear magneton.
$f_2^{\Sigma^\pm \Lambda}$ are given instead of $\rho_f^{\Sigma^\pm
\Lambda}$, since $f_1^{\Sigma^\pm \Lambda} = 0$}
\begin{tabular}{lcccccccc}
\hline
\\
Quantity & Experimental & CVC & RQM \cite{kell74} & MIT
 \cite{dono82} & LAPP \cite{lies87} & QCD \cite{cars88} &
 $\chi{\rm QSM_{br}}$ \cite{kimp97} & $\chi$QM\\
 & value\\
\hline
$\rho_f^{np}$ & $3.71 \pm 0.00$ (input) &
$3.71$ & $3.62$ & $3.63$ & $2.95$ & $3.71$ (input) & $3.16$ & $3.52$\\
$\rho_f^{\Sigma^- \Sigma^0}$ & - & $0.84$ & - & $1.35$ & - & - & $0.86$ &
 $1.30$\\
$f_2^{\Sigma^- \Lambda}$ & $3.52 \pm 3.52$ & $2.34$ & $2.67$ & $2.79$
 & $2.33$ & $2.48$ & $2.57$ & $2.73$\\
$f_2^{\Sigma^+ \Lambda}$ & - & $2.34$ & - & - & - & - & - & $2.72$\\
$\rho_f^{\Xi^- \Xi^0}$ & - & $-2.03$ & - & - & - & - & - & $-2.27$\\
\hline
$\rho_f^{\Sigma^- n}$ & $-1.78 \pm 0.61$ & $-2.03$ & $-1.95$ & $-2.04$
 & $-1.72$ & $-2.39$ & $-2.11$ & $-1.82$\\
 & $-1.71 \pm 0.27$ \protect\cite{hsue88}\\
$\rho_f^{\Xi^- \Sigma^0}$ & - & $3.71$ & - & $4.92$ & $3.33$ & $5.12$
 & $3.92$ & $3.84$\\
$\rho_f^{\Xi^- \Lambda}$ & $-0.44 \pm 0.46$ & $-0.12$ & - & $0.14$ &
 $-0.17$ & $0.16$ & $-0.33$ & $-0.06$\\
$\rho_f^{\Lambda p}$ & $2.43 \pm 1.49$ & $1.79$ & $1.98$ & $1.90$ &
$1.14$ & $2.44$ & $1.36$ & $1.38$\\
$\rho_f^{\Xi^0 \Sigma^+}$ & - & $3.71$ & - & - & - & - & - & $3.83$\\
\hline
\end{tabular}
\label{tab:rho}
\end{threeparttable}
\end{table*}

\begin{table*}
\begin{threeparttable}
\caption{The ratios $\rho_g^{BB'} \equiv
\frac{g_2^{BB'}}{g_1^{BB'}}$. The experimental value has been
obtained from \protect\cite{hsue88}}
\begin{tabular}{lccccccc}
\hline
\\
Quantity & Experimental & RQM \cite{kell74} & MIT \cite{dono82} &
LAPP \cite{lies87} & QCD \cite{cars88}
& $\chi{\rm QSM_{br}}$ \cite{kimp97} & $\chi$QM\\
 & value\\
\hline $\rho_g^{np}$ & - & $-0.45$ & $0$ & - & $0$ (input) & $0$ &
$0$ \tnote{a}
\\
$\rho_g^{\Sigma^- \Sigma^0}$ & - & - & $0$ & - & - & $0$ &
$0$ \tnote{a}
\\
$\rho_g^{\Sigma^- \Lambda}$ & - & $-0.46$ & $-0.03$ & $-0.12$
&
$-0.27$ & $-0.06$ & $-0.46$
\\
$\rho_g^{\Sigma^+ \Lambda}$ & - & - & $-0.03$ & - & - & - & $-0.41$
\\
$\rho_g^{\Xi^- \Xi^0}$ & - & - & - & - & - & - & $0$
\tnote{a}
\\
\hline
$\rho_g^{\Sigma^- n}$ & $-5.0 \pm 3.9$ & $-0.72$ & $0.27$ &
$-0.55$
& $-0.37$ & $0.12$ & $-0.14$
\\
$\rho_g^{\Xi^- \Sigma^0}$ & - & - & $0.42$ & $-0.25$
& $-0.03$ &
$0.18$ & $0.36$
\\
$\rho_g^{\Xi^- \Lambda}$ & - & - & $0.37$ & $-0.47$
& $-0.24$ &
$0.05$ & $0.05$
\\
$\rho_g^{\Lambda p}$ & - & $-0.66$ & $0.29$ & $-0.32$
& $-0.15$ &
$0.12$ & $0.12$
\\
$\rho_g^{\Xi^0 \Sigma^+}$ & - & - & - & - & - & - & $0.37$
\\
\hline
\end{tabular}
\begin{tablenotes}
\item[a]{The mass difference for baryons in the same isospin multiplet
has been neglected}
\end{tablenotes}
\label{tab:eta}
\end{threeparttable}
\end{table*}

\begin{table*}
\begin{threeparttable}
\caption{The weak form factors $g_{A}^{BB'}$ and
$g_{PT}^{BB'}$. The experimental values for $g_{A}^{BB'}$ have been
obtained from \protect\cite{barn96}, except for the ${g_1^{\Sigma^-
\Lambda}}$ and ${g_A^{\Xi^- \Sigma^0}}$ values, which
are CERN WA2 \protect\cite{bour82,gail84} results from branching
ratio measurements. The experimental values all assume that the weak
form factor $g_{2}=0$. The experimental value for $g_{PT}^{\Sigma^{-}n}$
has been obtained from \protect\cite{hsue88}}
\begin{tabular}{lcccc}
\hline
\\
Decay & \multicolumn{2}{c}{$g_{A}^{BB'}$} &
\multicolumn{2}{c}{$g_{PT}^{BB'}$} \\
 & Experimental value & $\chi$QM & Experimental value & $\chi$QM \\
\hline
$n \to p$ & $1.2601\pm 0.0025$ (average) & $1.12$ & - & $0$
\tnote{a} \\
$\Sigma^- \to \Sigma^0$ & - & $0.45$ & - & $0$ \tnote{a} \\
$\Sigma^- \to \Lambda $ \tnote{b} &
$(0.589\pm0.016$ \tnote{c} (CERN WA2)) & $0.55$ & - & $-0.25$ \\
$\Sigma^+ \to \Lambda $ \tnote{b} & - & $0.55$ & - & $0.22$ \\
$\Xi^- \to \Xi^0$ & - & $-0.22$ & - & $0$ \tnote{a} \\
\hline
$\Sigma^- \to n$ & $-0.20\pm 0.08$ \tnote{d} & $-0.22$ &
$1.00\pm0.66$ & $0.03$ \\
&$-0.340 \pm 0.017$ (average)& & & \\
$\Xi^- \to \Sigma^0$ & $(1.25 \pm 0.15$ \tnote{c} (CERN WA2)) &
$1.12$ & - & $0.40$\\
$\Xi^- \to \Lambda$ & $0.25\pm0.05$ (average) & $0.22$ & - & $0.01$ \\
$\Lambda \to p$ & $0.718\pm0.015$ (average) & $0.67$ & - & $0.08$ \\
$\Xi^0 \to\Sigma^+$ & - & $1.12$ & - & $0.41$ \\
\hline
\end{tabular}
\begin{tablenotes}
\item[a]{The mass difference for baryons in the same
isospin multiplet has been neglected}
\item[b]{$g_{1}^{\Sigma^{\pm}\Lambda}$ and
$g_{2}^{\Sigma^{\pm}\Lambda}$ are given instead of 
$g_{A}^{\Sigma^{\pm}\Lambda}$
and $g_{PT}^{\Sigma^{\pm}\Lambda}$, respectively, since
$f_{1}^{\Sigma^{\pm}\Lambda} = 0$}
\item[c]{Not listed in the Review of Particle Physics
\protect\cite{barn96}}
\item[d]{Evaluated using $g_{PT}^{\Sigma^-n}=1.00\pm0.66$
\cite{hsue88}}
\end{tablenotes}
\label{tab:gagpt}
\end{threeparttable}
\end{table*}

\end{document}